\documentclass[aps,prb,twocolumn,showpacs]{revtex4}

\usepackage{amsmath}
\usepackage{epsfig}
\newcommand{\bea}{\begin{eqnarray}}
\newcommand{\eea}{\end{eqnarray}}
\newcommand{\be}{\begin{equation}}
\newcommand{\ee}{\end{equation}}

\begin{document}

\title{Non-equilibrium charge noise and dephasing from  a spin-incoherent Luttinger liquid}

\author{Gregory A. Fiete$^{1,2}$ and Markus Kindermann$^{3,4}$}

\affiliation{$^1$Kavli Institute for Theoretical Physics, University of California, Santa Barbara, California 93106, USA\\$^2$Department of Physics, California Institute of Technology, MC 114-36, Pasadena, California 91125, USA\\$^3$Laboratory of Atomic and Solid State Physics, Cornell University, Ithaca, New York 14853-2501, USA\\ $^4$School of Physics, Georgia Institute of Technology, Atlanta, Georgia 30332, USA}

\begin{abstract}

We theoretically investigate the charge noise and dephasing in a metallic device in close proximity to a spin incoherent Luttinger liquid with a small but finite current. The frequency dependence of the charge noise exhibits a loss of frequency peaks corresponding to the $2k_F$ part of the density correlations in the electron liquid when the temperature $T$ is increased from values below the magnetic exchange energy $J$ of the electron gas to values above it.  The dephasing rate in a nearby metallic nanostructure also shows a   cross-over for $T \sim J$ and may exhibit a non-monotonic temperature dependence. For a range of temperatures the dephasing rate {\em decreases} with increasing temperatures.  The proposed experiments provide a convenient approach to probe the spin-incoherent Luttinger liquid and should be implementable in a wide variety of systems.

\end{abstract}

\date{\today}
\maketitle



\section{Introduction}

The noise of the electrical current has proven to be a powerful probe of small electronic structures.\cite{Beenakker:pt03, Blanter:prep00} 
While the equilibrium or Johnson-Nyquist noise only  contains information about the temperature and the linear response of a system,  the fluctuations out of equilibrium (such as in the presence of a voltage bias) carries a great wealth of information.  For example, the non-equilibrium current (shot) noise of a conductor reveals the granularity of electric charge and can be used to measure its fundamental value.  Such experiments have proved extremely useful in demonstrating convincingly that the $\nu =1/(2n+1)$ fractional quantum Hall effect states have a basic unit of charge  $e/(2n+1)$.\cite{Saminadayar:prl97,dePiccioto:nat97,Reznikov:nat99}  Measurements of the noise can also be used to determine the particle statistics of many-body systems with effects such as ``bunching" for bosons\cite{Brown:nat56} and ``anti-bunching" for fermions\cite{Henney:sci99,Oliver:sci99} predicted and observed.  
Finite-frequency noise measurements as discussed in this article have been proposed before to observe fractional charges in non-chiral Luttinger liquids \cite{trauzettel:prl04,lebedev:prb05}as well as many-body resonances in interacting nanostructures.\cite{sindel:prl05}

\begin{figure}[h]
\includegraphics[width=.95\linewidth,clip=]{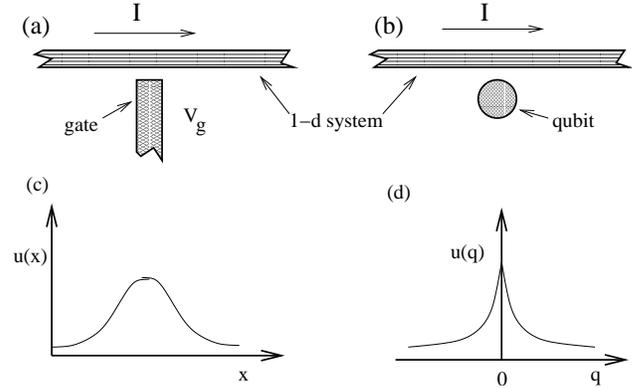}
\caption{\label{fig:schematic} Schematic of the proposed experiments described in the text and the dimensionless potential $u(x)$ at the gate/qubit due to a charge density $\delta \rho(x)$ at position $x$ along the 1-d system.  A current $I$ is assumed to flow in the 1-d system as indicated.  This could be a carbon nanotube or semiconductor quantum wire, for example.  In (a) a metallic gate is placed nearby and the fluctuations in the induced charge  $Q(t)=CV_g(t)=e\int dx u(x)\delta\rho(x,t)$ are measured. Here $e$ is the charge of the electron and $C$ is the capacitance.  In (b) a charge qubit is prepared in a quantum coherent state and its decoherence time is measured. In (c) $u(x)$  and its Fourier transform (d) are shown.  In order to observe the effects described in this paper, the size of the gate and qubit, and the distance to the 1-d system should be small compared to the inter-particle spacing in the 1-d system, $a$.  This ensures the Fourier components of the potential $u(2k_F)$ and $u(4k_F)$ and their corresponding contributions to noise/dephasing are not too small. }
\end{figure}

In this paper we study a schematic situation like that shown in Figs.~\ref{fig:schematic}a and \ref{fig:schematic}b.  A current $I$ is driven in a one dimensional (1-d) system such as a carbon nanotube or semiconductor quantum wire, and a small device such as a metallic gate or a charge qubit is placed in close proximity.\cite{Averin:prb94,Pedersen:prb98} Due to the discrete nature of the electrons (quasi-particles) in the 1-d system, there will be a time-dependent  charge $Q(t)$ induced on the gate in Fig.~\ref{fig:schematic}a and a time-dependent potential that dephases the charge states of the qubit in Fig.~\ref{fig:schematic}b.  We will address the frequency spectrum of the charge fluctuations $\langle Q(t)Q(0)\rangle$ on the gate in  Fig.~\ref{fig:schematic}a and the decoherence  rate of the qubit in Fig.~\ref{fig:schematic}b.  

We are interested in the particular situation where the interactions between electrons in the wire (we will refer to the 1-d system generically in this paper as a wire) are very strong and appreciable Wigner-crystal like correlations are present.  This typically means that $r_s=a/(2a_B)$ is large, where $a$ is the inter-particle spacing and $a_B$ is the Bohr radius for the material.  At large $r_s$ there can be an exponentially large separation of spin and charge energy scales.\cite{Matveev:prl04,Matveev:prb04}  At finite temperatures, this makes it possible to have highly excited spin degrees of freedom while keeping the charge degrees of freedom close to the ground state.  In 1-d this scenario is referred to as the ``spin incoherent Luttinger liquid'' (SILL).\cite{Fiete:rmp07} It may also be realized  at high electron densities in very thin wires. \cite{Fogler:prb05}

The SILL has received much attention recently because of its distinct properties that partially resemble those of a Luttinger liquid (LL) but partially do not.\cite{cheianov03,Cheianov04,Fiete:prl04,Fiete:prb05,Fiete_2:prb05,Fiete:prb06,Kindermann_noise:prb06,Kindermann:prl06,Kakashvili:cm06,Fiete:prl06,Matveev:prl06,Cheianov:cm04}  The crucial difference between the more familiar LL\cite{Voit:rpp95,Giamarchi,haldane81} and a SILL is the {\em combination} of (i) very strong particle-particle interaction and (ii) finite but small temperature.  As stated above, in a SILL the charge degrees of freedom are only very weakly influenced by the temperature, while the spin degrees of freedom are highly thermally excited.  In fact, below a critical temperature related to the fundamental energy scale in the spin sector, LL behavior is obtained in the same system exhibiting SILL behavior at higher temperatures.  This fact enables us to study the SILL within an effective low energy LL description as it is ``approached" by increasing the temperature from below the critical  value.  The details of this method of study are described below.  They have already proven useful in illuminating the electrical transport\cite{Fiete_2:prb05} and Coulomb drag\cite{Fiete:prb06} properties in SILL systems.  

The generic form of the single mode non-chiral (right and left movers present) 1-d Hamiltonian for energies small compared to the charge scale, but arbitrary compared to the spin scale is $H_{\rm elec}=H_c + H_s$, where
\be
\label{eq:H_c}
H_c=\hbar v_c \int \frac{dx}{2\pi}\left[\frac{1}{K_c} (\partial_x\theta_c(x))^2+K_c(\partial_x\phi_c(x))^2\right],
\ee
and
\be
\label{eq:H_s}
H_s=\sum_l  J_l {\vec S_l}\cdot {\vec S_{l+1}}.
\ee
The Hamiltonian \eqref{eq:H_c} describes the low-energy density fluctuations of the electron gas. The energy scale for the charge sector is set by $E_c \sim \hbar v_c/a$ where $\hbar$ is Planck's constant, $v_c$ is the collective charge mode velocity and $a$ is the average spacing between electrons.  The parameter $K_c$ describes the strength of the microscopic interactions and the bosonic fields appearing in Eq.~\eqref{eq:H_c} satisfy the commutation relations $[\partial_x\theta_c(x),\phi_c(x')]=i\pi \delta(x-x')$.  For strong interactions the spin degrees of freedom behave like a 1-d antiferromagnetic spin chain. In Eq.~\eqref{eq:H_s}, ${\vec S_l}$ is the spin of the $l^{th}$ electron,  and $J_l$ is the nearest neighbor exchange energy which depends on the local separation of electrons.  As discussed in Ref.[\onlinecite{Fiete:prb06}], to lowest order in local electron displacement from equilibrium, $J_l\approx J+J_1 (u_{l+1}-u_l)$ where $u_l$ is the displacement of the $l^{th}$ electron.  The main role of the term proportional to $J_1$ is to induce $2k_F$ oscillations in the density correlations after the higher energy density fluctuations of the electron gas have been integrated out.\cite{Fiete:prb06}  At the lowest energies, $T <J, E_c$, the effective density is given by\cite{Fiete:prb06}
\bea
 \label{eq:rho_eff_final}
\rho^{\rm eff}(x,t) = \rho_0 -\frac{\sqrt{2}}{\pi}\partial_x\theta_c(x,t)-\rho_0\left(\frac{J_1}{m\omega_0^2 a^2}\right)\nonumber \\ 
\times \sin\left(2k_Fx+\sqrt{2}\theta_c(x,t)\right)\sin(\sqrt{2}\theta_s(x,t))
\nonumber \\
+ \rho_0 \cos\left(4k_Fx+\sqrt{8}\theta_c(x,t)\right).\hspace{.5 cm}
\eea
At low energies $T<J$ the effective Hamiltonian of the SU(2) spin sector \eqref{eq:H_s} is\cite{Fiete:prb06}
\be
\label{eq:H_s_eff}
H^{\rm eff}_s=\hbar v_s \int \frac{dx}{2\pi}\left[\frac{1}{K_s}(\partial_x\theta_s(x))^2+K_s(\partial_x\phi_s(x))^2\right],
\ee
where $v_s\approx Ja/\hbar$, $K_s=1$ for $SU(2)$ symmetry, and the bosonic spin fields satisfy the same commutation relations as the charge fields and they commute with the charge fields.  In Eq.~\eqref{eq:rho_eff_final} the characteristic frequency of lattice oscillations $\omega_0$ is related to the charge velocity as $v_c=\omega_0a$. \cite{Fiete:prb06}  Note that in the effective description of the $2k_F$ part of the density oscillations \eqref{eq:rho_eff_final}, the amplitude is suppressed by a factor $\sim \left(\frac{J_1}{m\omega_0^2 a^2}\right)\ll 1$ relative to the corresponding value one would obtain from bosonizing a weakly interacting electron gas.\cite{Voit:rpp95}  

Returning to the main task of this paper, computing the charge fluctuation spectrum on the gate and the decoherence rate of the qubit,\cite{dephasing_LL_comment} we will find that both are determined by the Fourier transform of the density-density correlation function $\langle  (\rho^{\rm eff}(x,t) - \rho_0)(\rho^{\rm eff}(0,0) - \rho_0)\rangle$.  The key physical point is that when $J\ll T\ll E_c$, the $2k_F$ parts of this correlation function will be thermally washed out and the effects of losing these correlations will be observed in the charge fluctuation (noise) spectrum on the gate and the decoherence rate of the qubit.  In particular we find that the sharp gate response at the frequencies corresponding to the $2k_F$ oscillations vanishes, and the decoherence time of the qubit may exhibit non-monotonic behavior, including a region where the decoherence time (rate) increases (decreases) as the temperature increases, counter to the naive expectation.  See Fig.~\ref{fig:temp_dephasing}.

This paper is organized as follows.  In Sec.~\ref{sec:model} we introduce the basic formalism for calculating the non-equilibrium noise in a situation like that in Fig.~\ref{fig:schematic}.  In Sec.~\ref{sec:noise} we discuss the behavior of the noise  for strongly interacting electrons.  As the SILL state is approached from temperatures $T<J$ we describe the resulting cross-over in the noise spectrum when $T \sim J$.  In Sec.~\ref{sec:dephasing} we describe the formalism for computing the dephasing (decoherence) time $\tau_\varphi$ for a charge qubit in close proximity to a current carrying wire and evaluate  $\tau_\varphi$ as a function of temperature.  Finally, in Sec.~\ref{sec:conclusions} we describe the main conclusions of the work and prospects for experimental implementation.  

\section{Describing the fluctuations}
\label{sec:model}

In real physical systems the charge density is not uniform, so when current flows along a 1-d system as shown in Fig.~\ref{fig:schematic} time-dependent fluctuations in the potential outside the system are created.  If a metallic gate is nearby (Fig.~\ref{fig:schematic}a), these fluctuations will cause a time-dependent induced charge $Q(t)$ on it.\cite{Averin:prb94,Pedersen:prb98}  If a charge qubit is nearby (Fig.~\ref{fig:schematic}b) these same fluctuations will  lead to decoherence or dephasing.  Our main goal in this section is to lay out the relevant formalism for describing the non-equilibrium fluctuations on the gate and the qubit.\cite{pinning_comment}  

\begin{figure}[h]
\includegraphics[width=.95\linewidth,clip=]{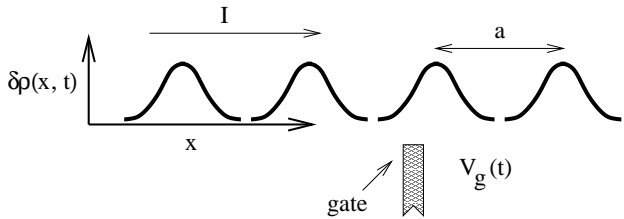}
\caption{\label{fig:solid_schematic} Schematic of a snapshot of the charge density $\delta \rho(x,t)$ in a strongly interacting 1-d system flowing by a metallic gate with induced voltage $V_g(t)=\frac{Q(t)}{C}=\frac{e}{C} \int dx u(x) \delta \rho(x,t)$.  (See Fig.~ \ref{fig:schematic}.)  The current $I\equiv ev_d/a$ in the 1-d system produces strong voltage and charge spikes on the gate with frequency $\omega_I=v_d(2\pi/a)=2\pi I/e$, corresponding to $4k_F\equiv 2\pi/a$ density modulations.  Here $v_d$ is the mean velocity of the charge $e$ due to the current $I$.  }
\end{figure}

Since the gate is assumed metallic, a net local charge imbalance in the wire, $e \delta \rho(x,t) dx$, will induce a charge $\delta Q(t)$ on the gate whose magnitude will depend on the proximity of the local charge in the wire to the gate.  We call the function describing the distance dependence $u(x)$, (see Fig.~\ref{fig:schematic}c for a schematic) and it has the same dependence on $x$ as the interaction potential between the charge $e \delta \rho(x,t) dx$ and $\delta Q(t)$.  The total charge on the gate is given by summing all the contributions along the wire,
 \be
 Q(t)=e\int dx u(x) \delta\rho(x,t).   
\ee
The charge fluctuations (noise) on the gate are thus determined by $\langle \{Q(t),Q(0)\}\rangle=e^2\int dx \int dx' u(x)u(x')\langle \{\delta \rho(x,t),\delta \rho(x',0)\}\rangle$, where $\{A,B\}=AB+BA$ and the expectation values are to be evaluated in the presence of a  finite current $I$. The frequency-dependent charge noise 
\be
S_Q(\omega)=\frac{1}{2}\int dt e^{i\omega t} \langle \{Q(t),Q(0)\}\rangle, 
\ee
can be expressed as
\be
\label{eq:noise}
S_Q(\omega)=\frac{e^2}{2\pi}\int dq |u(q)|^2 \frac{\chi(q,\omega)+\chi(q,-\omega)}{2},
\ee
where $\chi(q,\omega)$  is the Fourier transform of the density-density correlation function $\chi(x,t)=\langle \delta \rho(x,t)\delta \rho(0,0)\rangle $.  Here $\delta \rho(x,t)=\rho(x,t)-\rho_0$, with $\rho_0\equiv 1/a$, and $a$ is the inter-particle spacing.   Note that since $\langle \{Q(t),Q(0)\}\rangle$ is manifestly real and it only depends on $|t|$, $S_Q(\omega)=S_Q(|\omega|)$ is real.

When a current $I$ is flowing through the wire, the electrons are moving with a mean velocity $v_d=aI/e$. This velocity sets a characteristic frequency scale (discussed below) which will appear in the noise spectrum.  The density-density correlation function $\chi(q,\omega)$ appearing in \eqref{eq:noise} can be obtained from the equilibrium correlation function $\chi_0(q,\omega)$ by making a Galilean transformation so that \cite{Averin:prb94}
\be
\chi(x,t)=\chi_0(x-v_dt,t),
\ee
or equivalently,
\be
\label{eq:equil_relation}
\chi(q,\omega)=\chi_0(q,\omega-v_dq),
\ee
which can then be substituted into  \eqref{eq:noise}.  The problem of computing the non-equilibrium noise on the gate or the qubit is thus reduced to the problem of finding the {\em equilibrium} correlation function $\chi_0(q,\omega)$.  

Based on the structure of the expression for the effective density, Eq.~\eqref{eq:rho_eff_final}, we expect the Fourier transform of the density-density correlation function in the strong interaction limit (appropriate for realizing the SILL) to have the form
\be
\chi_0(q,\omega)\approx \chi_0^{q\approx 0}(q,\omega)+\chi_0^{2k_F}(q,\omega)+\chi_0^{4k_F}(q,\omega)\;,
\ee
plus higher order terms that are subdominant.  Each of these terms will lead to a characteristic frequency response in $S_Q(\omega)$ that can already be anticipated on purely physical grounds.
In the strong interaction Wigner crystal limit, the dominant periodicity of the density fluctuations are the $4k_F\equiv 2\pi/a$ pieces.  Thus, for a given current $I=ev_d/a$ there will be an electron passing by the gate with frequency $\omega_I\equiv v_d (2\pi/a) =2\pi I/e$.  (See Fig.~\ref{fig:solid_schematic}.) The $2k_F$ oscillations will pass by with 1/2 the frequency, $\omega_I/2$, and the $q\approx 0$ fluctuations will be peaked about zero frequency.   The central physics of our main results can now be seen:  When $T \gg J$ the $2k_F$ parts of the density-density correlation function will be washed out and this will cause the vanishing of the $\omega_I/2$ contribution to the noise and a loss of the $2k_F$ contribution to $\tau_\varphi$.  We now turn to a quantitative discussion of these points.

\section{Noise spectrum}
\label{sec:noise}

As we have seen in the previous section, the noise $S_Q(\omega)$ is determined by the equilibrium correlation function $\chi_0(q,\omega)$, and this has dominant low frequency parts coming from wavevectors near $q\approx 0, 2k_F$, and $4k_F$.  In this section we compute the various components of  $\chi_0(q,\omega)$ at finite temperature and use them to extract the frequency dependence of $S_Q(\omega)$.  We will focus on the case where the interactions are strong and the density fluctuations are described by \eqref{eq:rho_eff_final}.  The noise in the case of weak interactions and the case of strong interactions without the spin are discussed in Ref.[\onlinecite{Averin:prb94}], both at zero temperature.  Recent advances in computing the density-density correlation function for {\it weakly} interacting systems\cite{Pustilnik_2:prl06,Pustilnik:prl03} and for special forms of the interaction\cite{Pustilnik:prl06} may be used in \eqref{eq:noise} to compute the noise in these cases.

\subsection{The $\chi_0^{q\approx 0}(q,\omega)$ contribution to $S_Q(\omega)$}

Focusing first on the $q\approx 0$ components in \eqref{eq:rho_eff_final}, we compute the Fourier transform of $\chi_0^{q\approx 0}(x,t)= \langle \frac{\sqrt{2}}{\pi}\partial_x\theta_c(x,t)\frac{\sqrt{2}}{\pi}\partial_x\theta_c(0,0)\rangle$. Using the Fourier expansion 
\be 
\theta_c(x,t)=\frac{\pi}{\sqrt 2}\sum_q \sqrt{\frac{\hbar}{2 a L m \omega_q}} e^{iq x}(a_q e^{-i\omega_q t}+a^\dagger_{-q}e^{i\omega_qt}),
\ee
where $L$ is the length of the 1-d system, and $\omega_q=v_c |q|$ is the dispersion of the long wavelength density fluctuations.  The bosonic operators satisfy the commutation relations $[a_q,a^\dagger_{q'}]=\delta_{q,q'}$. Computing the Fourier transform of $\chi_0^{q\approx 0}(x,t)$ immediately leads to
\be
\label{eq:q0_FT}
\chi_0^{q\approx 0}(q,\omega)=\frac{q^2}{a^2}\frac{\pi \hbar L}{m\omega_q}\frac{\left[\delta(\omega-\omega_q)- \delta(\omega+\omega_q)\right]}{1-e^{-\beta \omega}},
\ee
which can then be substituted into  \eqref{eq:noise} using  \eqref{eq:equil_relation} to yield the noise contribution,
\bea
\label{eq:noise_0}
S_Q^{q\approx 0}(\omega)=\frac{e^2}{2\pi} \frac{\pi \hbar L|\omega|}{2 a^2mv_c}\Biggl[\frac{|u(\frac{\omega}{v_c-v_d})|^2 }{(v_c-v_d)^2}\hspace{2.7 cm}\nonumber \\
\times \left(\frac{1}{1-e^{-(1+\frac{v_d}{v_c-v_d})\beta |\omega|}}-\frac{1}{1-e^{(1+\frac{v_d}{v_c-v_d})\beta |\omega|}}\right)\nonumber \\
+ \frac{|u(\frac{\omega}{v_c+v_d})|^2}{(v_c+v_d)^2}\Biggl(\frac{1}{1-e^{-(1-\frac{v_d}{v_c+v_d})\beta |\omega|}}\hspace{2 cm}\nonumber \\
-\frac{1}{1-e^{(1-\frac{v_d}{v_c+v_d})\beta |\omega|}}\Biggr)\Biggr]. \nonumber \\
\eea
It is worth noting that $S_Q^{q\approx 0}(\omega) \sim |\omega|$ as $\omega \to 0$ for realistic forms of $u(x)$ which have finite $u(q=0)$.

 In the limit of a small current $I$ in the 1-d system (to prevent the system from being too far out of equilibrium) we expect the velocity $v_d$ to be much smaller than the charge velocity $v_c$.  Expanding the result \eqref{eq:noise_0} in $v_d/v_c$ yields current-dependent fluctuations only at  order  $(v_d/v_c)^2$.  These can be extracted by measuring $S_Q^{q\approx 0}(\omega)|_{I}-S_Q^{q\approx 0}(\omega)|_{I=0}\approx \frac{e^2}{2\pi} \frac{\pi \hbar L|\omega|}{2 a^2mv_c} |u(\frac{\omega}{v_c})|^2|\omega| 6 \left(\frac{v_d}{v_c}\right)^2.$

\subsection{The $\chi_0^{2k_F}(q,\omega)$ contribution to $S_Q(\omega)$}

The behavior of $\chi_0^{2k_F}(q,\omega)$ for temperatures $T <J$ and $T>J$ is the central contributing factor to the interesting temperature dependence of   $S_Q(\omega)$ and $\tau_\varphi$ when the 1-d system has interactions strong enough to produce a large separation in spin and charge velocities.  When $v_s \ll v_c$ and the system is at finite temperature, the spin-incoherent regime can be reached.  As we have emphasized in the introduction, the main effect in the density correlations at $T \gtrsim J$ is that the $2k_F$ components get washed out by thermal effects\cite{Fiete:prb06} and this effectively eliminates these contributions to the non-equilibrium noise and $\tau_\varphi$.   Let us first see how the $2k_F$ correlations are lost before we turn to the $T=0$ correlations and the corresponding noise.

From  \eqref{eq:rho_eff_final}, one sees that   $\chi_0^{2k_F}(q,\omega)$ for $0<T<J \approx \hbar v_s/a$ is given by the Fourier transform of  
\bea
\label{eq:rho_2kF_temp}
\langle \rho^{\rm eff}_{2k_F}(x,t)\rho^{\rm
  eff}_{2k_F}(0,0)\rangle=
\rho_0^2\left(\frac{J_1}{m\omega_0^2 a^2}\right)^2  \cos(2k_F x) \hspace{1 cm}\nonumber \\
\times\frac{(\pi
  Tr_c/v_c)^{K_c}}{\left[\sinh\left(\frac{\pi T}{v_c}(x-
    v_ct+ir_c)\right)\sinh\left(\frac{\pi T}{v_c}(x+
    v_ct-ir_c)\right)\right]^\frac{K_c}{2}}
\nonumber \\
\times\frac{(\pi Tr_s/v_s)^{K_s}}{\left[\sinh\left(\frac{\pi T}{v_s}(x-
    v_st+ir_s)\right)\sinh\left(\frac{\pi T}{v_s}(x+
    v_st-ir_s)\right)\right]^\frac{K_s}{2}},  \nonumber \\
\eea
where the infinitesimals $r_c,r_s={\cal O}(a)>0$.

In order to see how temperature affects the $2k_F$ correlations, it is instructive to consider the equal time correlation functions for $|x| > a,r_c,r_s$,
\bea
\label{eq:temp_sup}
\langle \rho^{\rm eff}_{2k_F}(x,0)\rho^{\rm
  eff}_{2k_F}(0,0)\rangle \approx
\rho_0^2\left(\frac{J_1}{m\omega_0^2 a^2}\right)^2  \cos(2k_F x) \hspace{1cm}\nonumber \\
\times\frac{(\pi
  Tr_c/v_c)^{K_c}}{\left[\sinh\left(\frac{\pi T}{v_c}x\right)\right]^{K_c}}
\frac{(\pi Tr_s/v_s)^{K_s}}{\left [\sinh\left(\frac{\pi T}{v_s}x\right)\right]^{K_s}},  \nonumber \\
< \rho_0^2\left(\frac{J_1}{m\omega_0^2 a^2}\right)^2 \!\!\! \left(\frac{\pi Tr_s}{v_s}\right)^{K_s}\!\!\!
\left(\frac{\pi Tr_c}{v_c}\right)^{K_c}\nonumber \\
\times e^{-c T/J}
 \frac{  \cos(2k_F x) }{\left[\sinh\left(\frac{\pi T}{v_c}x\right)\right]^{K_c}},\nonumber \\
\eea
where $c>0$ is a constant of order unity.  Eq.~\eqref{eq:temp_sup} shows clearly that when $T \gtrsim J$, the $2k_F$ density correlations are exponentially suppressed.  When $v_s \ll v_c$, one can take the zero temperature limit in the charge sector\cite{Fiete:prl04} (since one can simultaneously have $\pi T |x|/v_s \gg 1$ and  $\pi T |x|/v_c \ll 1$) and only an exponentially small error is made. This approximation gives,
\be 
\langle \rho^{\rm eff}_{2k_F}(x,0)\rho^{\rm eff}_{2k_F}(0,0)\rangle \lesssim  \left(\frac{T}{J}\right)^{K_s}\!\!\!\!e^{-c T/J} \frac{  \cos(2k_F x) }{\left(x/r_c\right)^{K_c}}.
\ee
Eq.\ (\ref{eq:temp_sup}) is not applicable to temperatures of the order of $J$ or larger because the bosonized description  of the spin excitations ceases to be valid. One can infer, however, from the solution of the original spin Hamiltonian Eq.\ (\ref{eq:H_s}), that the exponential decay continues at $T > J$.\cite{Subir}   

This exponential decay with temperature carries over to  the Fourier transform of the density-density correlation function at wavevector $2 k_F$, $\chi_0^{2 k_F}(q,\omega)$.

  We turn now to 
the zero temperature limit of $\chi_0^{2 k_F}$.
To do so, we must compute the Fourier transform 
\begin{widetext}
\be
\label{eq:2kF_integral_start}
\bar{\chi}_0^{2k_F}(q_\pm,\omega)\approx \rho_0^2 \left(\frac{J_1}{m\omega_0^2 a^2}\right)^2 \int_{-\infty}^{\infty}dx \int_{-\infty}^\infty dt e^{i(\omega t -q_\pm x)}\frac{r_c^{K_c}}{[(x-v_ct+ir_c)(x+v_ct-ir_c)]^{\frac{K_c}{2}}}\frac{r_s^{K_s}}{[(x-v_st+ir_s)(x+v_st-ir_s)]^{\frac{K_s}{2}}},
\ee
\end{widetext}
where $q_\pm=q\pm 2k_F$, and $\chi_0^{2k_F}(q,\omega)=\bar{\chi}_0^{2k_F}(q_+,\omega)+\bar{\chi}_0^{2k_F}(q_-,\omega)$.  From \eqref{eq:2kF_integral_start} one can see that the singularity near $x=v_ct$ yields a contribution $\sim |\omega - v_c q_\pm|^{K_c/2+K_s-1}$ and the singularity near $x=v_st$ yields a contribution $\sim |\omega - v_s q_\pm|^{K_c+K_s/2-1}$.  The singularities at $x=-v_ct,-v_st$ have the same behavior only with $q_\pm \to -q_\pm$.  For $SU(2)$ symmetry $K_s=1$, and one sees the contribution from the spin singularity,  $\omega \approx \pm v_s q_\pm$, leads to a divergence in $\bar{\chi}_0^{2k_F}(q,\omega)$ only for $K_c<1/2$.  Otherwise, it has an inverted cusp form.  On the other hand, the contribution from the charge singularity, $\omega \approx \pm v_c q_\pm$, always has a cusp form.   This is shown in schematic form in figure 8 of the review article by Voit\cite{Voit:rpp95}, but there appears to be a typo in the exponent for $\omega \approx \pm v_c q_\pm$.

The zero temperature analytical properties of $\bar{\chi}_0^{2k_F}(q_\pm,\omega)$ can be extracted by making the change of variables $z=x+v_st$ and $\bar z=x-v_st$ and defining 
\be
\eta \equiv v_s/v_c <1,
\ee
which gives
\begin{widetext}
\bea
\label{eq:2kF_integral_second}
\bar{\chi}_0^{2k_F}(q_\pm,\omega)\approx \frac{\rho_0^2}{2v_s} \left(\frac{J_1}{m\omega_0^2 a^2}\right)^2 \int_{-\infty}^{\infty}\!\!\!\!\!d\bar z \frac{e^{-i\omega \bar z/v_s}}{(\bar z +i r_s)^{\frac{K_s}{2}}} \int_{-\infty}^\infty \!\!\!\!\!dx \frac{e^{i(\omega  -v_s q_\pm) x/v_s} r_s^{K_s}}{(2x-\bar z -i r_s)^{\frac{K_s}{2}}} \hspace{4 cm}\nonumber \\
\times
\frac{r_c^{K_c}}{[(x(1-\eta^{-1})+\eta^{-1}\bar z+ir_c)(x(1+\eta^{-1})-\eta^{-1}\bar z-ir_c)]^{\frac{K_c}{2}}}.
\eea
\end{widetext}
Note that when the outer integral over $\bar z$  is taken all the singularities  lie in the lower half-plane and therefore the result is proportional to $\theta(\omega)$.  If we further eliminate $x$ in favor of $\bar z$ and $z$, we find
\begin{widetext}
\bea
\label{eq:2kF_integral_third}
\bar{\chi}_0^{2k_F}(q_\pm,\omega)\approx \frac{\rho_0^2}{2v_s} \left(\frac{J_1}{m\omega_0^2 a^2}\right)^2 \int_{-\infty}^{\infty}\!\!\!\!\!d\bar z \frac{e^{-i(\omega +v_s q_\pm)\bar z/(2 v_s)}r_s^{K_s/2}}{(\bar z +i r_s)^{\frac{K_s}{2}}} \int_{-\infty}^\infty \!\!\!\!\!dz \frac{e^{i(\omega  -v_s q_\pm) z/(2v_s)} r_s^{K_s/2}}{(z -i r_s)^{\frac{K_s}{2}}} \hspace{4 cm}\nonumber \\
\times
\frac{r_c^{K_c}}{[((1-\eta^{-1})z+(1+\eta^{-1})\bar z+i2r_c)((1+\eta^{-1})z+(1-\eta^{-1})\bar z-i2r_c)]^{\frac{K_c}{2}}}.
\eea
\end{widetext}
Since $\eta^{-1}>1$, when the $z$ integral is taken, all the singularities lie in the upper half-plane giving a result proportional to $\theta(\omega -v_s q_\pm)$ and when the $\bar z$ integral is taken all the singularities lie in the lower have plane giving a result proportional to $\theta(\omega +v_s q_\pm)$.  Combining these results, we find $\bar{\chi}_0^{2k_F}(q_\pm,\omega) \propto \theta(\omega)\theta(\omega^2 -(v_s q_\pm)^2)$.  
 For repulsive interactions, that is for $K_c<1$, $K_s \approx 1$, all singularities in Eq.\ (\ref{eq:2kF_integral_third}) have the form $1/z^\alpha$ with $\alpha <1$. The integrals thus converge and  after having analysed the analytic properties of  (\ref{eq:2kF_integral_third}), resulting in the factor $\theta(\omega)\theta(\omega^2 -(v_s q_\pm)^2)$, the limit $r_c,r_s\to 0$ may be taken, 
 \begin{widetext}
\bea
\label{eq:2kF_integral_fourth}
\bar{\chi}_0^{2k_F}(q_\pm,\omega)\propto \theta(\omega)\theta(\omega^2 -(v_s q_\pm)^2) \int_{-\infty}^{\infty}\!\!\!\!\!d\bar z \frac{e^{-i(\omega +v_s q_\pm)\bar z/(2 v_s)}}{\bar z ^{\frac{K_s}{2}}} \int_{-\infty}^\infty \!\!\!\!\!dz \frac{e^{i(\omega  -v_s q_\pm) z/(2v_s)} }{z ^{\frac{K_s}{2}}} \hspace{4 cm}\nonumber \\
\times
\frac{1}{[((1-\eta^{-1})z+(1+\eta^{-1})\bar z)((1+\eta^{-1})z+(1-\eta^{-1})\bar z)]^{\frac{K_c}{2}}}.
\eea
\end{widetext}
To obtain $S_Q$ we first evaluate $S_Q^{2k_F(1-)}(\omega)$, the $q_-$ contribution to $\chi^{2 k_F}$ in the first term of Eq.\ (\ref{eq:noise}),
 \begin{equation}
S_Q^{2k_F(1-)}(\omega) =\frac{e^2}{2\pi}\frac{1}{2}\int dq |u(q)|^2 \bar{\chi}_0^{2k_F}(q-2k_F,\omega-v_dq).
\end{equation}
 The theta functions in Eq.\ (\ref{eq:2kF_integral_fourth}) constrain the $q$ integration to $q_{\rm min} < q <q_{\rm max}$ where $q_{\rm min}=\frac{\omega-v_s2k_F}{v_d-v_s}$ and $q_{\rm max}=\frac{\omega+v_s2k_F}{v_s+v_c}$.  For small currents, we expect $v_d <v_s$, so that we can expand in the ratio $v_d/v_s$.  Thus, $q_{\rm min}\approx -\frac{\omega}{v_s}+2k_F +\frac{\omega_I/2}{v_s}$ and  $q_{\rm max}\approx \frac{\omega}{v_s}+2k_F -\frac{\omega_I/2}{v_s}$, where $\omega_I\equiv v_d (2\pi/a)$ was defined at the end of Sec.~\ref{sec:model}.  Shifting $q$ by $2k_F$ then gives
\bea
\label{eq:2k_F_noise_initial}
S_Q^{2k_F(1-)}(\omega) =\frac{e^2}{2\pi}\frac{1}{2}\int_{-\left(\frac{\omega}{v_s}-\frac{\omega_I/2}{v_s}\right)}^{\frac{\omega}{v_s}-\frac{\omega_I/2}{v_s}} dq\hspace{2cm}\nonumber \\
\times |u(q+2k_F)|^2 \bar{\chi}_0^{2k_F}(q,\omega-\omega_I/2-v_dq).
\eea
We now extract the asymptotic behavior of $S_Q$ near $\omega \approx \omega_I/2$. For this we assume that $(\omega-\omega_I)/2v_s \ll 2 k_F$. If $u(q)$ is analytic at $q=2 k_F$, which we assume, we may for this replace $u(q)$ by $u(2k_F)$ in Eq.\ (\ref{eq:2k_F_noise_initial}). By scaling the integration variables by $q \to q/(\omega-\omega_I/2)$ and $z,\bar{z} \to z/(\omega-\omega_I/2),\bar{z}/(\omega-\omega_I/2)$ in \eqref{eq:2kF_integral_fourth} the asymptotic scaling of the resulting integral may then be extracted:
\be
S_Q^{2k_F(1-)}(\omega) \sim  \left(\frac{J_1}{m\omega_0^2 a^2}\right)^2 |u(2k_F)|^2 |\omega - \omega_I/2|^{K_s+K_c-1}.
\ee
 Carrying through the same calculation for the $\bar{\chi}_0^{2k_F}(q+2k_F,\omega-v_dq)$ contribution one finds,
\be
S_Q^{2k_F(1+)}(\omega) \sim \left(\frac{J_1}{m\omega_0^2 a^2}\right)^2 |u(2k_F)|^2 |\omega + \omega_I/2|^{K_s+K_c-1}.
\ee
Likewise, the contributions from the second term in Eq.\ (\ref{eq:noise})   take on a similar form and we finally arrive at the zero temperature result
\be 
\label{eq:2k_F_noise}
S_Q^{2k_F}(\omega) \sim \left(\frac{J_1}{m\omega_0^2 a^2}\right)^2\sum_\pm  |u(2k_F)|^2 \left|\omega \pm \frac{\omega_I}{2}\right|^{K_s+K_c-1}.
\ee

Eq.~\eqref{eq:2k_F_noise} shows that indeed the $2k_F$ contributions to the noise are centered about $\pm \omega_I/2$ with a power-law behavior that depends on the interaction parameters of the 1-d system.

\subsection{The $\chi_0^{4k_F}(q,\omega)$ contribution to $S_Q(\omega)$}

The next dominant  component of the noise comes from the $4k_F$ part of the density correlations \eqref{eq:rho_eff_final},
\bea
\label{eq:rho_4kF_temp}
\langle \rho^{\rm eff}_{4k_F}(x,t)\rho^{\rm eff}_{4k_F}(0,0)\rangle=
\rho_0^2 \cos(4k_F x)\hspace{3 cm}\nonumber \\
\times\frac{(\pi T\alpha/v_c)^{4K_c}}{\left[\sinh\left(\frac{\pi T}{v_c}(x- v_ct+ir_c)\right)\sinh\left(\frac{\pi T}{v_c}(x+ v_ct-ir_c)\right)\right]^{2K_c}}.\nonumber \\
\eea
Taking the zero temperature limit of this correlation function and evaluating the Fourier transform after making the coordinate transformation $z=x+v_ct$, $\bar z = x-v_ct$ one finds
\bea
\chi_0^{4k_F}(q_\pm,\omega) \approx \theta(\omega)\theta(\omega^2 -(v_c q_\pm)^2)(\omega^2-(v_cq_\pm)^2)^{2K_c-1}\nonumber \\
\times \frac{\rho_0^2}{v_c}r_c^{4K_c} \Gamma(1-2K_c)^2 \sin(2\pi K_c)^2,\;\;\;\;
\eea
where again we have $\chi_0^{4k_F}(q,\omega)=\bar \chi_0^{4k_F}(q_+,\omega)+\bar \chi_0^{4k_F}(q_-,\omega)$.  After the Galilean shift \eqref{eq:equil_relation}   we find
\be
S_Q^{4k_F}(\omega)\sim \sum_\pm |u(4k_F)|^2  |\omega \pm \omega_I|^{4K_c-1},
\ee
which exhibits frequency dependence centered around $\pm \omega_I$.

Collecting the $q\approx 0, 2k_F$ and $4k_F$ contributions we have,
\be
S_Q(\omega)= S_Q^{q\approx 0}(\omega)+S_Q^{2k_F}(\omega)+S_Q^{4k_F}(\omega),
\ee
with
\bea
S_Q^{q\approx 0}(\omega)&\propto&|u(0)|^2 |\omega|,\nonumber\\
S_Q^{2k_F}(\omega)&\propto&\left(\frac{J_1}{m\omega_0^2 a^2}\right)^2\sum_\pm |u(2k_F)|^2 \left|\omega \pm \frac{\omega_I}{2}\right|^{K_s+K_c-1}\nonumber\\
S_Q^{4k_F}(\omega)&\propto& \sum_\pm |u(4k_F)|^2|\omega \pm \omega_I|^{4K_c-1}.
\eea
The frequency dependence of the charge noise measured at a gate nearby a quantum wire thus displays power law singularities at the frequencies $\omega_I/2$ and $\omega_I$ that are observable at low  temperatures, $kT \ll \omega, \omega_I$. The singularity $S_Q^{2k_F}$ at $\omega\approx \omega_I/2$, however, becomes  exponentially small as $T \gtrsim J$.

\section{Dephasing}
\label{sec:dephasing}

Having discussed the finite frequency noise in a metallic gate like the one in Fig.~\ref{fig:schematic}a, we now turn to the situation shown in Fig.~\ref{fig:schematic}b.  As we will see immediately below, the dephasing rate $\tau_\varphi$ of the charge qubit depicted there is determined by the zero frequency noise produced by the nearby quantum wire.\cite{Buttiker:prb00,Pilgram:prl02}   We assume an effective coupling Hamiltonian between wire and qubit of the form
\be
H_{wq}= \sigma_z \left(\frac{e^2}{2C}\right)\int{dx\, u(x)\delta\rho(x)},
\ee
where $\sigma_z$ is the third Pauli matrix acting on the space of qubit states. The dephasing rate $\tau_\varphi$ of the qubit is then determined by the zero frequency fluctuations of $\delta \rho$,
\be
\label{eq:tau_phi_def}
\frac{1}{\tau_\varphi}=\frac{1}{2}\int_{-\infty}^\infty dt K(t),
\ee
where 
\be
\label{eq:K_t}
K(t) = \left(\frac{e^2}{2C}\right)^2\int dxdx'\, u(x) u(x') \langle \delta \rho(x,t) \delta\rho(x',0) \rangle, 
\ee
so that
\bea
\label{eq:tau_phi_final}
\frac{1}{\tau_\varphi}&=& \frac{1}{2}\left(\frac{e^2}{2C}\right)^2\int{\frac{dq}{2\pi} |u(q)|^2 \chi(q,\omega=0)} \nonumber \\
&=&\frac{1}{2}\left(\frac{e^2}{2C}\right)^2\int \frac{dq}{2\pi} |u(q)|^2 \chi_0(q,-v_d q).
\eea
As we have seen earlier, $\chi_0(q,\omega)\approx \chi_0^{q\approx 0}(q,\omega)+\chi_0^{2k_F}(q,\omega)+\chi_0^{4k_F}(q,\omega)$, plus higher order terms that are subdominant.  In the expression for the dephasing rate $1/\tau_\varphi$ each of these terms is multiplied by the appropriate factor of $|u(q)|^2$.  As shown in Fig.~\ref{fig:schematic}d $u(q)$ typically  montonically decreases with increasing $|q|$.  Thus, the smaller the momenta, the larger the contribution from $\chi_0(q,\omega)$.   

It turns out (somewhat reminiscently of the Coulomb drag case) that the $q\approx 0$ piece does not contribute to the dephasing due to the same phase space restrictions imposed by the linear dispersion that kill this contribution to the drag between quantum wires.\cite{Fiete:prb06}  This is easily seen from \eqref{eq:q0_FT} which shows that $\chi_0^{q\approx 0}(q,-v_d q) \equiv 0$ since $v_c\neq v_d$ ensuring the delta function is zero for all finite $q$.  This leaves the two dominant contributions to $\tau_\varphi$ coming from the $2k_F$ and $4k_F$ pieces,
\bea 
\label{eq:tau_phi_approx}
\frac{1}{\tau_\varphi} \approx \frac{1}{\tau_\varphi^{2k_F}}+\frac{1}{\tau_\varphi^{4k_F}}.
\eea
We obtain the temperature dependence of each of these pieces from Eqs.\ \eqref{eq:rho_2kF_temp} and \eqref{eq:rho_4kF_temp} together with  \eqref{eq:tau_phi_final}. We now assume small currents such that $kT \gg \omega_I$. We find
\be
 \frac{1}{\tau_\varphi^{2k_F}} \sim |u(2k_F)|^2\rho_0^2 \left(\frac{J_1}{m\omega_0^2a^2}\right)^2T^{K_s+K_c-1}e^{-c T/J},
\ee
and
\be
 \frac{1}{\tau_\varphi^{4k_F}} \sim |u(4k_F)|^2\rho_0^2 T^{4K_c -1}.
\ee

\begin{figure}[t]
\includegraphics[width=.95\linewidth,clip=]{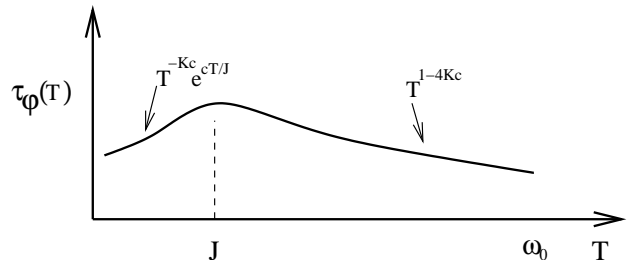}
\caption{\label{fig:temp_dephasing} Schematic of the temperature dependence of the dephasing time of a qubit   as described in the text.   The geometry of the proposed experiment is shown in Fig.~\ref{fig:schematic}.  The temperature dependence of the dephasing time $\tau_\varphi$ is given by Eq.~\eqref{eq:tau_phi_final}, and an approximate form by Eq.~\eqref{eq:tau_phi_approx}.  The constant $c$ in the exponential is ${\cal O}(1)$ and $\omega_0$ is a high frequency cut off in the charge sector. The figure assumes $1/4 <K_c <1$ and $K_s=1$.} 
\end{figure}

The temperature dependence of $\tau_\varphi$ is shown schematically in Fig.~\ref{fig:temp_dephasing} for $1/4 <K_c <1$.  In this regime, the dephasing time may exhibit non-monotonic behavior with temperature in a way somewhat reminiscent of the Coulomb drag resistance between two parallel quantum wires.\cite{Fiete:prb06}  For $T \lesssim J$, the dephasing time may actually {\rm increase} as the temperature increases.  This is counter to the naive expectation that heating up the system (bath) is likely to {\em decrease} the coherence time because presumably the fluctuations are increasing.  In fact, the fluctuations the qubit experiences decrease as $T \to J$ from below because the $2k_F$ fluctuations that dominate the noise (assuming $|u(2k_F)|^2 \left(\frac{J_1}{m\omega_0^2a^2}\right)^2> |u(4k_F)|^2$) in this temperature range are washed out for $T \gtrsim J$ leaving only the weaker $4k_F$ contributions.   An observation of this behavior is strong evidence for the spin-incoherent Luttinger liquid.

\section{Conclusions}
\label{sec:conclusions}

We have studied the non-equilibrium noise spectrum and the dephasing rate in a small device in close proximity to a spin-incoherent Luttinger liquid with finite current $I$.  Both the noise, $S_Q(\omega)$, and the dephasing time, $\tau_\varphi$, are sensitive to the density correlations in the strongly interacting 1-d system, and these correlations qualitatively change when the temperature $T$ becomes of order the magnetic exchange energy $J$.  In the noise, this leads to a loss in the frequency response near $\omega \approx \omega_I/2\equiv \pi I/e$, corresponding to a loss of the $2k_F$ component of the density correlations.  In the dephasing rate, a non-monotonic dependence on temperature should be observed  with a transition around $T\sim J$, again due to the loss of the $2k_F$ component of the density correlations. 

So far, there are experimental indications pointing to the realization of the SILL in quantum wires with low electron density,\cite{Steinberg:prb06} but these data are only preliminary and restricted to relatively few particle numbers.    At present extremely high quality quantum wires with low electron density appear to be a very promising candidate for observing spin-incoherent effects and there are now many falsifiable theoretical predictions to be put to the test.  We hope those detailed here will soon be tested experimentally.

\acknowledgments

We thank L. Balents for enlightening discussions. This work was supported by  NSF Grant numbers PHY99-07949, DMR04-57440, DMR03-34499, and the Packard Foundation. G.A.F. was also supported by the Lee A. DuBridge Foundation.   We gratefully acknowledge the hospitality of the Aspen Center for Physics where part of this work was carried out.

\appendix
\section{Derivation of dephasing rate for a quantum dot}

A problem completely analogous to the dephasing of a qubit  as discussed in Sec~\ref{sec:dephasing} is that of dephasing of a quantum dot. Following Levinson\cite{Levinson:prb00,Levinson:epl97} we outline here how one arrives in that case at the expressions for the dephasing rate used in section \ref{sec:dephasing}.
We assume the Hamiltonian for the problem is given by $H=H_{\rm elec}+H_{\rm QD}+H_{\rm int}$
where $H_{\rm elec}$ is given by Eqs.\eqref{eq:H_c} and \eqref{eq:H_s},
\be
H_{\rm QD}=\epsilon_0c^\dagger c,
\ee
and
\be
\label{eq:int}
H_{\rm int}=c^\dagger c \left(\frac{e^2}{2C}\right)\int dx u(x)\delta \rho(x),
\ee
where as before $e$ is the charge of the electron and $C$ is the capacitance of the quantum dot. From Eq.~\eqref{eq:int} it is clear that the effect of the interaction of the electrons in the wire with the quantum dot is to shift the level $\epsilon_0$ of the dot by $\frac{e^2}{2C}\int dx u(x)\delta \rho(x)$.  Indeed, if $\delta \rho(x)$ depends on time, then the quantum dot level will  ``jiggle" and this will lead to dephasing, or decoherence.\cite{Fiete:pra03,Stern:pra90}  Below we define and calculate the dephasing rate $\tau_\varphi$ as a function of temperature.

It is conceptually convenient to discuss the decoherence in terms of density matricies.  We will assume the quantum dot has been initially prepared\cite{Sprinzak:prl00} in a state $|\psi\rangle= \alpha_0 |0\rangle + \alpha_1 |1\rangle$ with $|1\rangle \equiv c^\dagger |0\rangle $.  This leads to the density matrix
\be
\hat \rho = |\psi\rangle \langle \psi |.
\ee
Coherence in a quantum system is manifest in non-zero off diagonal elements of the density matrix.  At time $t=0$ we have $\langle c(0) \rangle = \alpha_1 \alpha_0^* \neq 0$, so the initial state is coherent.
We wish to compute the decay of the off-diagonal density matrix elements (from density fluctuations in the 1-d system) given by $\langle c(t)\rangle$, where $c(t)=e^{iHt}ce^{-iHt}$.  By taking the time derivative of this equation one finds $\frac{d}{dt} c(t)=-i(\epsilon_0+W(t))c(t)$, where  $W(t)=e^{iHt}We^{-iHt}$ and $W\equiv \frac{e^2}{2C}\int dx u(x) \delta\rho(x)$. This can be integrated to yield
\be
c(t) = c(0)e^{-i\epsilon_0t}T_te^{-i\int_0^t dt' W(t')}.
\ee
Therefore, we obtain the {\em exact} expression\cite{Levinson:epl97}
\be \label{A5}
\langle c(t)\rangle=e^{-i\epsilon_0t}\langle c(0) T_te^{-i\int_0^t dt'  W(t')}\rangle.
\ee
We now assume that the interaction $u(x)$ between the quantum dot and the 1-d system is sufficiently weak that there are negligible back-action effects of the dot on the 1-d system.  We thus approximate the time evolution of $W(t) \approx e^{i( H_{\rm elec}+H_{\rm QD})t}  W e^{-i(H_{\rm elec}+H_{\rm QD})t}=e^{i H_{\rm elec}t} W e^{-iH_{\rm elec}t}$, effectively keeping the lowest order corrections in $W$.  In a cumulant expansion of (\ref{A5}) the leading contribution is
\be
\langle c(t)\rangle \approx \langle c(0)\rangle e^{-i\epsilon_0t} e^{-\Phi(t)},
\ee
with
$\Phi(t)=\frac{1}{2}\int dt_1 \int dt_2 K(t_1-t_2)$ and  $K(t_1-t_2)=\langle W(t_1) W(t_2)\rangle$, where we have used the fact that the correlator is a function only of the time difference.  By making the change of variables $y_1=t_1-t_2$ and $y_2=(t_1+t_2)/2$, we can express $\Phi(t)$ as
\be
\Phi(t)=t \frac{1}{2} \int_{-t}^t dy_1 K(y_1).
\ee
The function $K(t)$ has some characteristic cut-off time $\tau_c$ such that $K(t)\approx 0$ when $t>\tau_c$.  Thus, for long times the integral above can be extended to infinity defining a dephasing time $\tau_\varphi$,
\be
\langle c(t)\rangle \approx \langle c(0)\rangle e^{-i\epsilon_0t}e^{-t/\tau_\varphi}
\ee
with
\be
\frac{1}{\tau_\varphi}=\frac{1}{2}\int_{-\infty}^\infty dt K(t).
\ee



\begin{thebibliography}{46}
\bibitem[{pin()}]{pinning_comment}
\bibinfo{note}{Throughout this paper we neglect any back action of the gate or
  dot on the quantum wire. In principle, the gate and dot can provide a local
  potential disturbance in the wire that can pin the Wigner crystal at
  sufficiently small source-drain voltages. However, estimating the energy
  scale for this is not straightforward at it relies on a renormalization group
  analysis as discussed in Ref.[\onlinecite{Giamarchi}].}
\end{thebibliography}

\end{document}